# A Probabilistic Approach for Authenticating Text or Graphical Passwords Using Back Propagation

**ASN Chakravarthy**[†]
Associate Professor, Dept.of CSE&IT
Sri Sai Aditya Institute of Science Technology,
Suram Palem, E.G.Dist, Andhra Pradesh, India

**Prof.P S Avadhani**[††],
Professor, Dept. of CS & SE,
Andhra University,
Visakhapatnam Dist, Andhra Pradesh, India

**Abstract**
Password authentication is a common approach to the system security and it is also a very important procedure to gain access to user resources. In the conventional password authentication methods a server has to authenticate the legitimate user. In our proposed method users can freely choose their passwords from a defined character set or they can use a graphical image as password and that input will be normalized. Neural networks have been used recently for password authentication in order to overcome pitfall of traditional password authentication methods. In this paper we proposed a method for password authentication using alphanumeric password and graphical password. We used Back Propagation algorithm for both alphanumeric (Text) and graphical password by which the level of security can be enhanced. This paper along with test results show that converting user password in to Probabilistic values enhances the security of the system

## 1. Introduction

Computer security has become a very important part of human life. Recently authentication has become an important issue among many access control mechanisms. Secure networks allows only intended recipient to intercept and read a message addressed to him. Thus protection of information is required against possible violations than compromise its secrecy [1]. Secrecy is compromised if information is disclosed to users not authorized to access it. Password authentication is one of the mechanisms that are widely used to authenticate an authorized user [2].

The main limitation in using the traditional password authentication method is that, a server must maintain a password table that stores each user's ID and password. Therefore, the server requires extra memory space to store the password table. The table is shown in Table.1. When a user logs into a computer, he types in the ID and password. The server searches the password table and checks if the password is legal. However, this method is dangerous. The password information table could be read or altered by an intruder. An intruder can also append a new ID and password into the table.

| USER NAME | PASSWORD |
|---|---|
| Harsh | 25-may1991 |
| Vamsi | Vss123 |
| Suresh | Abecke |
| Aditya | Letmein |
| Sanjay | 24dk03k |

Table 1. Password table

| USER NAME | ENCRYPTED PASSWORD |
|---|---|
| Harsha | ↑ǨNʒ⊥⌋ |
| Vamsi | kJɋ26v |
| Suresh | DPts£8 |
| Aditya | ¥~   ¥2p |
| Sanjay | efeaeolg |

Table 2. Verification table

Password table is protected using hash functions later and instead of password table [3] verification table containing hashed password (encrypted) [4] will be stored in the server. Even though password is encrypted still there is a chance of modification of the verification table since it is open access environment.

There are many disadvantages in using this type of approach.
  i. Attacker can easily change the details of users by using attacks like SQL-Injection.
  ii. Password table occupies a lot of memory**.**

To avoid this problem we proposed a password authentication method using Back Propagation algorithm for both alphanumeric password (textual) and graphical password.

Before discussing the actual method we briefly discuss the related schemes for password authentication. Following the related review, our proposed scheme is presented. The implementation method, experiment, observations and results, advantages and disadvantages, conclusion and future scope are discussed in the last section of this paper

## 2. Related review
**Authentication:**
Authentication is the act of confirming the truth of an attribute of a datum or entity. This might involve confirming the identity of a person, tracing the origins of an artifact, ensuring that a product is what it's packaging and labeling claims to be, or assuring that a computer program is a trusted one.

One familiar use of authentication and authorization is access control. A computer system that is supposed to be used only by those authorized must attempt to detect and exclude the unauthorized. Access to it is therefore usually controlled by insisting on an authentication procedure to establish with some degree of confidence the identity of the user, hence granting those privileges as may be authorized to that identity.

Common examples of access control involving authentication include:
- A captcha is a means of asserting that a user is a human being and not a computer program.
- A computer program using a blind credential to authenticate to another program
- Entering a country with a passport
- Logging in to a computer
- Using a confirmation E-mail to verify ownership of an e-mail address
- Using an Internet banking system
- Withdrawing cash from an ATM

### 2.1 Password

Using strong passwords lowers overall risk of a security breach, but strong passwords do not replace the need for other effective security controls. The effectiveness of a password of a given strength is strongly determined by the design and implementation of the authentication system software, particularly how frequently password guesses can be tested by an attacker and how securely information on user passwords is stored and transmitted. Risks are also posed by several means of breaching computer security which are unrelated to password strength.

#### 2.1.1 Alphanumeric Password (textual)

Alphanumeric password is derived from a Character Set. There are so many types of Character sets depending upon the application where we need authentication. One of the well known Character Set is the American Standard Code for Information Interchange (ASCII). It is a character-encoding scheme based on the ordering of the English alphabet.

ASCII includes definitions for 128 characters: 33 are non-printing control characters (now mostly obsolete) that affect how text and space is processed; 94 are printable characters, and the space is considered as an invisible graphic.

A common attack against password authenticated systems is the dictionary attack. An attacker can write a program that, imitating a legitimate user, repeatedly tries different passwords, say from a dictionary, until it gets the correct password. We present an alternative defense against dictionary attacks by using Graphical password.

#### 2.1.2 Graphical Password

The most common computer authentication method is to use alphanumerical usernames and passwords. This method has been shown to have significant drawbacks. For example, users tend to pick passwords that can be easily guessed. On the other hand, if a password is hard to guess, then it is often hard to remember. To address this problem, we a developed authentication method that use pictures as passwords [5].

Use graphics (images) instead of alphanumerical passwords
- A picture is worth a thousand words
- Humans remember pictures better than words

Although the main argument for graphical passwords is that people are better at memorizing graphical passwords than text-based passwords, the existing user studies are very limited and there is not yet convincing evidence to support this argument. Our preliminary analysis suggests that it is more difficult to break graphical passwords using the traditional attack methods such as brute force search, dictionary attack, or spyware. However, since there is not yet wide deployment of graphical password systems, the vulnerabilities of graphical passwords are still not fully understood.

### 2.2 Password Authentication

The idea of password assignment is to base the authentication of an identity on something the user knows. In other words, the distinguishing characteristic is knowledge. In a security perspective it should be seen as a user-remembered key. Passwords should ideally be a random string of letters, numbers and other symbols. Unfortunately that is far from reality in most systems. The whole notation of passwords is based on an oxymoron. The idea is to have a random string that is easy to remember.

Drawbacks with traditional password authentication [6]
- User password difficult to memorize.
- User cannot freely choose is password
- User cannot change his password
- It cannot with stand forgery attack

Our proposed method can with stand Replay and forgery attack

### 2.3 Neural Networks (NN)

Neurons are the most basic unit. Neurons are interconnected. These connections are not equal, as each connection has a connection weight. Groups of networks come together to form layers. These weights are what give the neural network the ability to recognize certain patterns. Adjust the weights; the neural network will recognize a different pattern.

Adjustment of these weights is a very important operation. The process of training is adjusting the individual weights between each of the individual neurons until we achieve close to the desired output [7].

The essence of neural networks is to learn the behavior of actors in the system (E.g. Users, daemons etc). Neural networks use its learning algorithms to learn about the relationship between input and output vectors and to generalize them to extract new input/output relationships. The advantage of using neural networks over statistics is its ease in expressing and learning nonlinear relationships between variables [8]. Experiments using neural networks resulted that the behavior of super user (roots) are predictable.

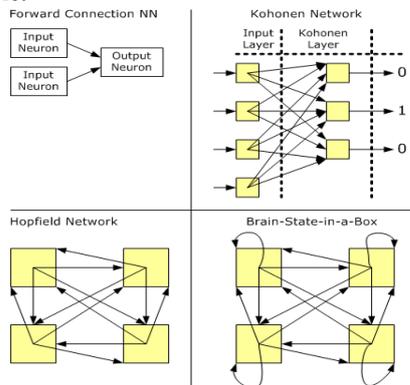

Fig. 1. Some examples of neural networks

One of the popular NN is called the Back Propagation neural network (BP NN) which will be discussed next.

This BP NN consists of three layers:
1. Input layer with three neurons.
2. Hidden layer with two neurons.
3. Output layer with two neurons.

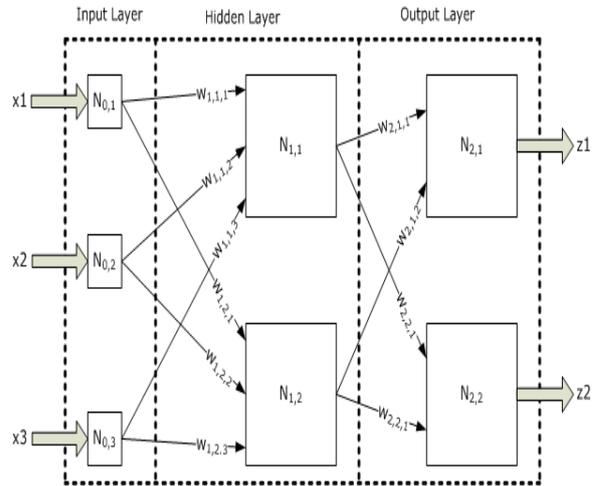

Fig. 2. Back Propagation

**Note that:**

1. The output of a neuron in a layer goes to all neurons in the following layer.
2. Each neuron has its own input weights.
3. The weights for the input layer are assumed to be 1 for each input. In other words, input values are not changed.
4. The output of the NN is reached by applying input values to the input layer, passing the output of each neuron to the following layer as input.
5. The Back Propagation NN must have at least an input layer and an output layer. It could have zero or more hidden layers.

The number of neurons in the input layer depends on the number of possible inputs we have, while the number of neurons in the output layer depends on the number of desired outputs. The number of hidden layers and how many neurons in each hidden layer cannot be well defined in advance, and could change per network configuration and type of data. In general the addition of a hidden layer could allow the network to learn more complex patterns, but at the same time decreases its performance. You could start a network configuration using a single hidden layer, and add more hidden layers if you notice that the network is not learning as well as you like. Not surprisingly, researchers have also tried to use neural networks in Cryptography. A recent survey of the literature indicates that there has been an increasing interest in the application of different classes of neural networks to problems related to cryptography in the past few years. Recent works have examined the use of neural networks in cryptosystems. Typical examples include key management, generation and exchange protocols; visual cryptography; pseudo random generators; digital watermarking; and steganalysis [9].

## 3. Proposed Scheme:

To overcome the disadvantages of traditional password authentication schemes, we introduced a novel technique for password authentication. In this technique there is no need of storing usernames and passwords in the server. We used a feed forward neural network and we will **train** it using usernames as input and password as output. When a particular user submits his login credentials we have given his username as input to network and we checked whether the output of network and specified password are equal or not, if both are equal the user is an authorized person.

### 3.1 Method

1. Define an own character set for alphanumeric data including special characters or we can use ASCII/GRAY CODE/EBCDIC/UNICODE.
2. Normalize each character in to probabilistic values in the range [0,1]

$$C_N = \frac{C_{take} - C_{min}}{C_{max} - C_{min}} \quad (1)$$

where, $C_{take}$ = Character taken,
$C_{max}$ = maximum value of character set,
$C_{min}$ = minimum value of character set

3. a) The Normalized password data is supplied as input to a multi feed forward Back Propagation neural network with one or more hidden layers. This produces encoded password in real values within the range [0,1].
   b) De normalize encoded data in to character notation (for memorization and backup)
4. a) Guessing password data is given as input for Back Propagation algorithm with one or more hidden layers to produce decrypted data.
   b) If it is matched the password is authenticated, otherwise it is invalid.
5. for de normalization use

$$C_{take} = C_N (C_{max} - C_{min}) \quad (2)$$

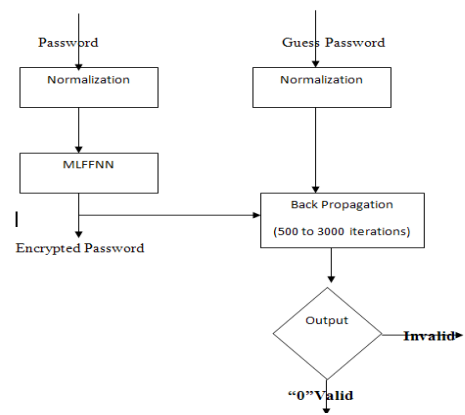

Fig. 3. Proposed Scheme.

### 3.1 Probabilistic Approach Using Feed Forward Neural Network:

In feed forward network the output of neurons (unit) in one layer will be passed as input to the next layer and this process continues until the output layer units gets an input from previous layers. Finally these output units yield an output. The output of network depends on Input, Connection strengths (Weight values), and Output function used in each layer. If we modify any of the above the output of the network will be changed. By taking this fact as an advantage we can perform encryption so that no attacker can decrypt it easily.

Here if 'P' is a row matrix representing input and 'W' is a matrix representing weights of the network then a feed forward network produces cipher text in the following way.

$$C_1 = \sum_{j=0}^{n} P_i W_{ij} \quad (3)$$

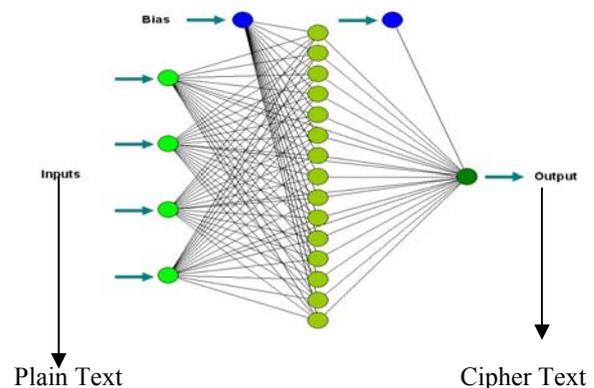

Figure 4. Feed forward Network

## 4. Experiment
**Context Setting:**
After performing the training we stored all weight values in a weight table as shown below. Whenever a user submits his login credentials network uses this weight values to produce output.

| W11 | W12 | W13 | W14 | W15 |
|---|---|---|---|---|
| W21 | W22 | W23 | W24 | W25 |
| W31 | W32 | W33 | W34 | W35 |
| W41 | W42 | W43 | W44 | W45 |
| W51 | W52 | W53 | W54 | W55 |

Table 3. Weights table

In order to define a character set we use following things.
- We can change order of the characters in a particular Character Set
- We can set the Maximum and Minimum values for the Character Set

Any organization which wants to use this novel password authentication technique can define their own Character Set by changing the order of the characters in the Character Set and giving their own maximum and minimum values for the Character Set. If the organization wants to use existing character sets like ASCII, UNICODE etc., still they can use our technique and in order to increase security they can change the order of characters, even they can change constant assigned to each character.

| Character | Unique code | Character | Unique code | Character | Unique code |
|---|---|---|---|---|---|
| A | 11 | J | 20 | S | 29 |
| B | 12 | K | 21 | T | 30 |
| C | 13 | L | 22 | U | 31 |
| D | 14 | M | 23 | V | 32 |
| E | 15 | N | 24 | W | 33 |
| F | 16 | O | 25 | X | 34 |
| G | 17 | P | 26 | Y | 35 |
| H | 18 | Q | 27 | Z | 36 |
| I | 19 | R | 28 | | |

Table 4. Example Character Set (text)

Here minimum value = 11 and maximum value = 36

In the normalization we will convert each unique number assigned to a character in to **probabilistic value**.

| Character | Probabilistic value | Character | Probabilistic value | Character | Probabilistic value |
|---|---|---|---|---|---|
| A | 0.00 | J | 0.36 | S | 0.72 |
| B | 0.04 | K | 0.40 | T | 0.76 |
| C | 0.08 | L | 0.44 | U | 0.80 |
| D | 0.12 | M | 0.48 | V | 0.84 |
| E | 0.16 | N | 0.52 | W | 0.88 |
| F | 0.20 | O | 0.56 | X | 0.92 |
| G | 0.24 | P | 0.60 | Y | 0.96 |
| H | 0.28 | Q | 0.64 | Z | 1.00 |
| I | 0.32 | R | 0.68 | | |

Table 5. Normalized Character Set

In our proposed method of password authentication we use two types of passwords
- Text password – alphanumerical Character Set
- Graphical password – predefined images or stored pictures

Here we can use any one of the above as password and we can train the neural network so that it can authenticate users

### 4.1 Text password
Here we will convert the password into its corresponding probabilistic values and we can use those values as input to the neural network.

| Password | Normalized Password |
|---|---|
| LETMEIN | 0.44#0.16#0.76#0.48#0.16#0.32#0.52 |
| APPLE | 0#0.6#0.6#0.44#0.16# |
| GETIN | 0.24#0.16#0.76#0.32#0.52 |

Here we used '#' as delimiter to separate each probabilistic value

Table 6. Normalized values for text password

The main problem with the alphanumeric passwords is that once a password has been chosen and learned the user must be able to recall it to log in. But, people regularly forget their passwords. If a password is not frequently used it will be even more susceptible to forgetting.

The recent surveys have shown that users select short, simple passwords that are easily guessable, for example, personal names of their family members, names of pets, date of birth etc [10].

### 4.2 Graphical password

Graphical password schemes have been proposed as a possible alternative to text-based schemes, motivated partially by the fact that humans can remember pictures better than text; psychological studies supports such assumption [11]

Before giving the image as password the image should be converted in to its RGB values and these values will be normalized using our normalization function. we can't give image directly as input to the neural network. So here we converted image into matrix (or text).

### 4.2.1 Conversion of image to matrix (or text)

By using the below procedure we can convert any image into a matrix consisting of set of numbers representing all the pixels of the image.

After converting image into a matrix consisting of set of numbers we can give it to the neural network as input and we can train it using username and image matrix as a training sample.

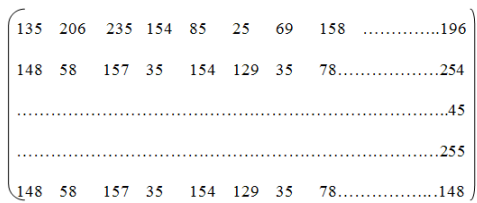

Fig. 5. Matrix representation an image

Read color of each pixel of the image. Convert the color into red, green and blue (RGB) parts as each color can be produced using these colors.

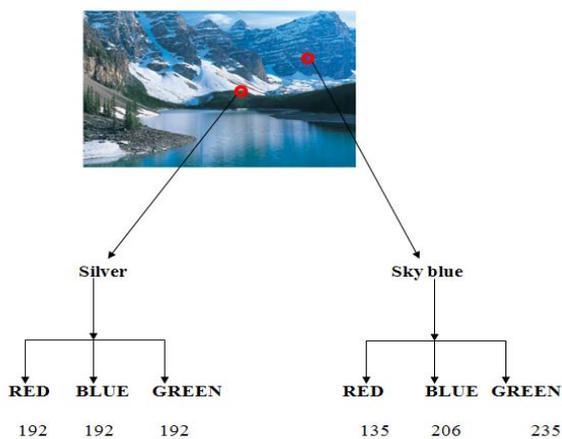

Fig. 6. Reading pixel values

Here we can normalize numbers in the matrix to increase security. As Red, Blue, Green can have maximum of 255 and minimum of 0 we can use following formula to normalize the numbers of matrix.

$$C_N = \frac{C_t}{255} \quad C_t \rightarrow \text{Value of Red or Green or Blue component} \quad (4)$$

Whenever new users are creating accounts, the network has to adjust weights so that it can recognize all the users who are registered. This process of changing weights is called learning. Here in order to do learning we used **back propagation** algorithm.

### 4.3 Implementation details

The neural network layers are implemented as arrays of structures. The nodes of the layers are implemented as follows:

```
[Serializable]
struct PreInput
{
    public double Value;
    public double[] Weights;
};

[Serializable]
struct Input
{
    public double InputSum;
    public double Output;
    public double Error;
    public double[] Weights;
};

[Serializable]
struct Hidden
{
    public double InputSum;
    public double Output;
    public double Error;
    public double[] Weights;
};

[Serializable]
struct Output<T> where T : IComparable<T>
{
    public double InputSum;
    public double output;
    public double Error;
    public double Target;
    public T Value;
};
```

**The layers are implemented as follows :**

```
private PreInput[ ] PreInputLayer;
private Input[ ] InputLayer;
private Hidden[ ] HiddenLayer;
private Output<string>[ ] OutputLayer;
```
**Training the network:**

- Apply normalized input to the neural network.
- Calculate the output in the form of normalized values in the range [0...1].
- Compare the resulting output with the desired output for the given input. This is called the error.
- Modify the weights for all neurons using the error.
- Repeat the process until the error reaches an acceptable value (e.g. error < 1%), which means that the neural network was trained successfully, or if we reach a maximum count of iterations, which means that the neural network training was not successful.

**It is represented as shown below:**
```
void TrainNetwork(TrainingSet,MaxError)
{
   while(CurrentError>MaxError)
   {
      foreach(Pattern in TrainingSet)
      {
         ForwardPropagate(Pattern);//calculate output
         BackPropagate()//fix errors, update weights
      }
   }
}
```

**Back Propagation:**

```
private void BackPropagate()
{
   int i, j;
   double total;
   //Fix Hidden Layer's Error
   for (i = 0; i < HiddenNum; i++)
   {
      total = 0.0;
      for (j = 0; j < OutputNum; j++)
      {
         total += HiddenLayer[i].Weights[j] *
                  OutputLayer[j].Error;
      }
      HiddenLayer[i].Error = total;
   }
   //Fix Input Layer's Error
   for (i = 0; i < InputNum; i++)
   {
      total = 0.0;
      for (j = 0; j < HiddenNum; j++)
      {
         total += InputLayer[i].Weights[j] *
                  HiddenLayer[j].Error;
      }
      InputLayer[i].Error = total;
   }
   //Update The First Layer's Weights
   for (i = 0; i < InputNum; i++)
   {
      for(j = 0; j < PreInputNum; j++)
      {
         PreInputLayer[j].Weights[i] += LearningRate
                         * InputLayer[i].Error
                         * PreInputLayer[j].Value;
      }
   }
   //Update The Second Layer's Weights
   for (i = 0; i < HiddenNum; i++)
   {
      for (j = 0; j < InputNum; j++)
      {
         InputLayer[j].Weights[i]+=LearningRat
                         * HiddenLayer[i].Error
                         * InputLayer[j].Output;
      }
   }
   //Update The Third Layer's Weights
   for (i = 0; i < OutputNum; i++)
   {
      for (j = 0; j < HiddenNum; j++)
      {
         HiddenLayer[j].Weights[i] +=
            LearningRate  *  OutputLayer[i].Error  *  HiddenLayer[j].Output;
      }
   }
}
```

## 5. Results and Observations

In back propagation method we will calculate error at the output layer and it will be propagated to the previous hidden layers and then to the input layer. Basing on the error at all the layers weights will be adjusted to get the correct output or to decrease the error rate.

### 5.1 For Textual Passwords

This screen indicates the Maximum Error and Maximum error that we may ignore while training the network.Error can be decreased either by changing learning rate parameter or by changing number of input and hidden units.Number of patterns are restricted by number of output units , if we want to recognize more patterns we have to use more output units.

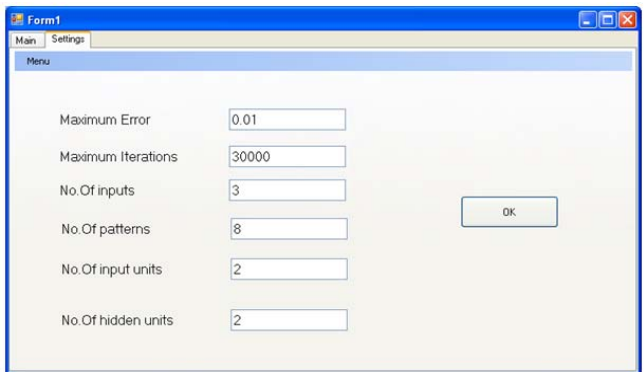

Fig. 7. Screen showing how to setup network and backpropagation algorithm

### 5.1.1 Training the Network

This scheme used back propagation algorithm to train the network.Since backpropagation method is very slow ,it may take more time to train the network.Once the training is completed it will displays the following message.

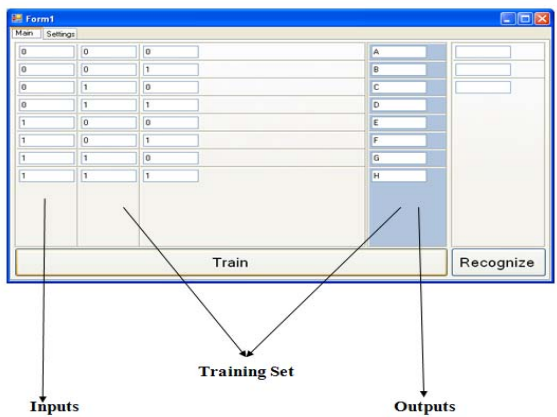

Fig. 8. Screen showing training the network using BP

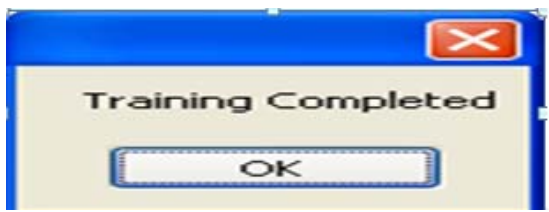

Fig.9 .Training completed message

### 5.1.2 Checking the user authorization

When the user enters the password it will be matched with trained password . Our scheme just brings the weights of the password and compare with given password weights and validates the user if the error is less.

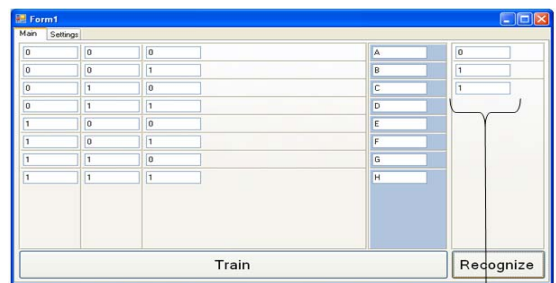

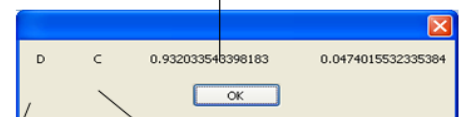

Fig.10.Validating the user

### 5.2 For Graphical Passwords

Maximum error indicates in the fig.11 is the maximum error that we may ignore in the training.

We can decrease error by doing following things
- By changing learning rate parameter
- By changing number of input and hidden units

Number of patterns are restricted by number of output units , if we want to recognize more patterns we have to use more output units. Here we have to specify directory which contains patterns.

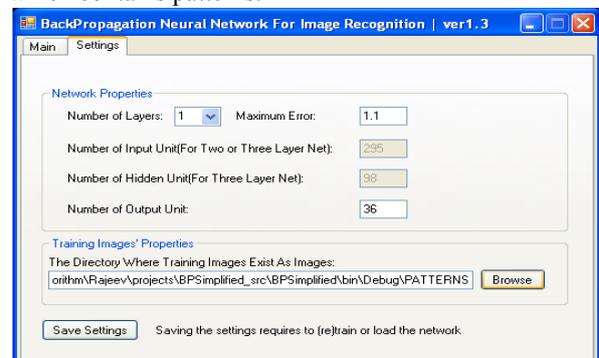

Fig.11. Screen showing how to setup network and backpropagation algorithm

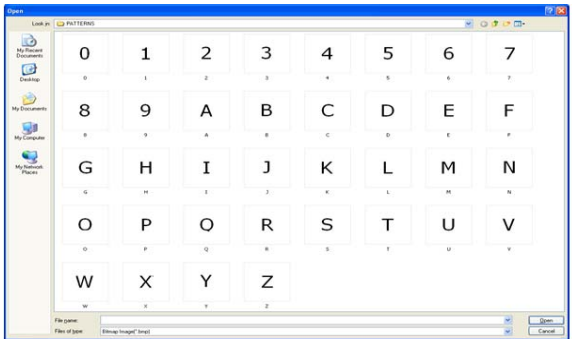
Fig.12. Screen shot Showing Patterns

### 5.2.1 Training the Network

Here after clicking the "Train Network" button it will load all the images located in the specified directory and it will adjust the weights.

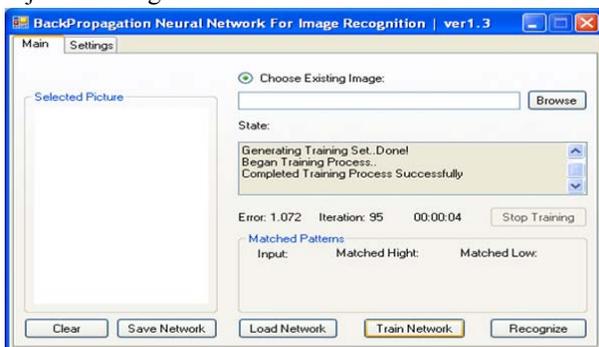
Fig.13. Training process

### 5.2.2 Checking whether a user is authorized or not

Here we have to select image to be tested by clicking the "Browse" button. After clicking "Recognize" button it takes an image as input and try to recognize it , if it can recognize the user is authorized.
Here let us consider a pattern that matched with less error.

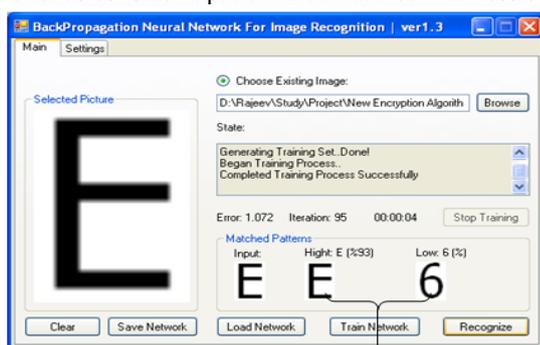
Fig.14. User validation

## 6. Advantages:

i. This method is very difficult to attack (Provides More Security).To decrypt cipher text attacker has to identify weight matrix (Even one element of matrix changes attacker can't decrypt), No. of Hidden layers, Output function, Character set (Including order of Characters in Character set), Minimum and Maximum values used in character set. Even though we use existing character set we can improve security by changing order of characters or minimum value to change unique number and probabilistic values associated to each character so that attacker may confuse in guessing the unique numbers or probabilistic values. We can increase security by increasing no. of Hidden layers. The users (Organization) of this algorithm can define their own character set, by doing so users can add new characters into their character set.

## 7. Disadvantages:

Since the output of this BPNN method is in the form of probabilistic values, the system can introduces noise, due which we may not do the efficient authentication.
Training time for BPNN is extremely large. Easy to remember passwords are vulnerable to password guessing attacks

## 8. Conclusion

In this paper, authentication using back propagation is implemented for both textual and graphical passwords. In the training process normalized input values were used for enhancing the password authentication. The past decade has seen a growing interest in using graphical passwords as an alternative to the traditional text-based passwords. In this scheme, the server does not store or maintain password or verification table. The server only stores the weights of the classification network.The system users can freely choose their password and the servers is required to retain only the pair user ID and password. The password authentication scheme can prevent the replay attack; the intruder cannot obtain a login password through the open network and replay the password to login to a server. Server only stores the weights of the network [5]. Our method prevents Replay attack.
      When all the networks are given with same training set, each one spends different amount of times to adjust their weight values. In all the cases bidirectional associative memory spends less amount of time comparing with all other networks and feed forward network spends huge amount of time as it uses back propagation for learning Table 7 shows these details more clearly.

| Training Time | | | |
|---|---|---|---|
| **BPNN** | 360 | 450 | 500 |
| **HPNN** | 136 | 50 | 100 |
| **BSB** | 30 | 49 | 80 |
| **BAM** | 25 | 49 | 70 |

Table 7. Training times of different networks for different inputs

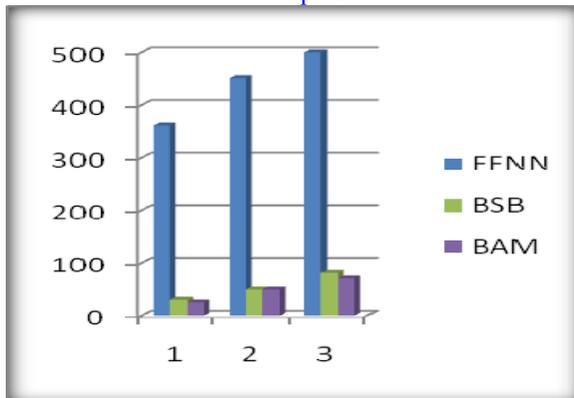

Fig. 15. Training times of different networks for different inputs

## 9. Future Work

We can also provide a virtual keypad through which password can be entered and can define some special characters in the character set for text passwords. For graphical passwords we can draw images or symbols on the virtual screen and can use those images as passwords. We want to enhance the method for other associative memories like Bidirectional memories, Hopfield Networks, Brain in state Box. Also we want to enhance the BPNN method for hand written Graphical passwords and hand written signatures.

## Acknowledgments

The authors wish to thank many anonymous referees for their suggestions to improve this paper.

## References

1) Khalil Shihab," A Backpropagation Neural Network for Computer Network Security", Journal of Computer Science 2 (9): 710-715, 2006.
2) Li-Hua Li, Iuon-Chang Lin, and Min-Shiang Hwang, " A Remote Password Authentication Scheme for Multiserver Architecture Using Neural Networks", IEEE Transactions On Neural Networks, VOL. 12, NO. 6, November 2001.
3) G. Horng, "Password authentication without using password table," Inform. Processing Lett., vol. 55, pp. 247–250, 1995.
4) M. Udi, "A simple scheme to make passwords based on one-way function much harder to crack," Computer Security, vol. 15, no. 2, pp. 171–176, 1996.
5) Robert Biddle, Sonia Chiasson, P.C. van Oorschot ," Graphical Passwords: Learning from the First Twelve Years", January 4, 2011. Technical Report TR-11-01, School of Computer Science, Carleton University, 2011.
6) Wei-Chi Ku," Weaknesses and Drawbacks of a Password Authentication Scheme Using Neural Networks for Multiserver Architecture", IEEE TRANSACTIONS ON NEURAL NETWORKS, VOL. 16, NO. 4, JULY 2005.
7) M. S. Obaidat and D. T. Macchiarolo, "An on-line neural-network system for computer access security," IEEE Trans. Ind. Electron., vol. 40, pp. 235–242, 1993.
8) "A multilayer neural-network system for computer access security," IEEE Trans. Syst., Man, Cybern., vol. 24, pp. 806–813,May 1994.
9) T. Schmidt, H. Rahnama , A. Sadeghian, "A Review Of Applications Of Artificial Neural Networks In Cryptosystems",*Seventh International Symposium on Neural Networks*, June 6-9, *2010* Shanghai, China.
10) G. Agarwal and R.S. Shukla," Security Analysis of Graphical Passwords over the Alphanumeric Passwords ", Int. J. Pure Appl. Sci. Technol., 2010.
11) Xiaoyuan Suo, Ying Zhu, and G. Scott Owen, "Graphical Passwords: A Survey", Proceedings of 21st Annual Computer Security Applications Conference, December 5-9, 2005, Tucson, Arizona.

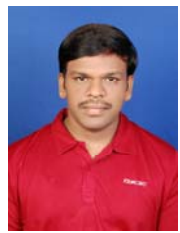

[†]**A.S.N Chakravarthy** received his M.Tech (CSE) from JNTU, Anantapur , Andhra Pradesh, India. Presently he is working as an Associate Professor in Computer Science and Engineering in Sri Sai Aditya Institute of Science Technology, SuramPalem, E.G.Dist, AP, India. His research area includes Network Security, Cryptography, Intrusion Detection, Neural networks.

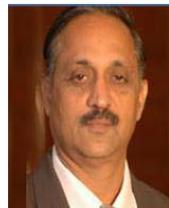

[††]**Prof. P.S.Avadhani** did his Masters Degree and PhD from IIT, Kanpur. He is presently working as Professor in Dept. of Computer Science and Systems Engineering in Andhra University college of Engg., in Visakhapatnam. He has more than 50 papers published in various National / International journals and conferences. His research areas include Cryptography, Data Security, Algorithms, and Computer Graphics, Digital Forensics and Cyber Security.

o0o